\documentclass[12pt]{article}
\usepackage[cp850]{inputenc}
\usepackage{amssymb,amsmath}
\usepackage{english}
\title{Symmetric and asymmetric nuclear matter in the relativistic approach
at finite temperatures}
\author{H. Huber, F. Weber, and M. K. Weigel\\Sektion Physik, Universit\"at
M\"unchen\\
Am Coulombwall 1, D-85748 Garching, Germany}

\begin{document}

\baselineskip15pt
\maketitle

\vskip1cm
PACs numbers:  21.65.+f, 21.60.Jz, 24.10.Cn, 97.60\,Jd

\begin{abstract}
The properties of hot matter are studied in the frame of the relativistic
Brueckner-Hartree-Fock theory. The equations are solved self-consistently in
the full Dirac space.
For the interaction we used the potentials given by Brockmann and Machleidt.
The obtained critical temperatures are smaller than in most of the
nonrelativistic investigations. We also calculated the thermodynamic
properties of hot matter in the relativistic Hartree--Fock approximation,
where the force parameters were adjusted to the outcome of the relativistic
Brueckner--Hartree--Fock calculations at zero temperature. Here, one obtains
higher critical temperatures, which are comparable with other relativistic
calculations in the Hartree scheme.
\end{abstract}

\newpage
\section{Introduction}
The properties of hot and dense nuclear matter play an essential role in the
understanding of high-energy heavy-ion collisions, supernova explosions and
proto-neutron stars. For that reason the problem of hot nuclear matter has
been studied over the last decades in several investigations, which however
were predominantly performed within the nonrelativistic
scheme \cite{1a}, using either effective density dependent interactions
\cite{1a,1,2,23} or the Brueckner approach \cite{1a,3,4,5,5a}. In the
relativistic approach investigations of the equations of state for $T\neq 0$
are relatively scarce. The majority of such calculations were performed in
the relativistic Hartree approximation (RH), where the extension to finite
temperatures is straightforward. Details of this scheme are given, for
instance, in Refs.\,\cite{23}, \cite{6}-\cite{11}. More complicated are the
relativistic Hartree--Fock-- \cite{13,11} and the Brueckner--Hartree--Fock
approximation \cite{12}, and the application to finite nuclei \cite{14}.

 In this contribution we will concentrate on the relativistic
Brueckner--Hartree--Fock treatment (RBHF) of symmetric and asymmetric nuclear
matter generalizing the formalism as described in Refs.\,\cite{15,15a} to
$T\neq 0$. The RBHF--approach seems to be of special interest, since it is
known for $T = 0$ that the resulting EOSs are much stiffer than their
nonrelativistic counterparts \cite{15b}. To our knowledge such an
investigation has been only performed so far by the Groningen group for
symmetric nuclear matter \cite{12}. As described in more details in
Refs.\,\cite{12,15}, their method uses a nonunique ansatz for the T--matrix
in terms of five Fermi invariants, which can lead to ambiguous results for
the self-energies (see, e.g., Refs.\,\cite{17,18}). In order to avoid this
problem we solve, according to the original scheme of the Brooklyn group
\cite{19}, the RBHF--approximation in the full Dirac space, which is more
tedious (for a more detailed discussion, see Ref.\,\cite{15}). Since the
formal structure of the problem is the same as for $T=0$, where one has to
solve three coupled equations, namely the Dyson equation for the one-body
Green's function $G$, the (reduced) Bethe--Salpeter equation for the
effective scattering matrix $T$ in matter and the equation for the
self--energy $\Sigma$, we will not repeat here the equations. As in
Refs.\,\cite{12,15,15a} we will restrict ourselves to the incorporation of
intermediate positive--energy nucleon states only, where now the Fermi step
functions are replaced by the Fermi distribution functions $n_{\vec p}\,(T)$
at finite  temperature $T$ (for details, see Ref.\,\cite{21}).  The Green's
function obeys for $T\neq 0$ the spectral representation \cite{11,20}
\begin{equation}
\label{I.1}
G(p) = \int d\omega\,A(\vec{p},\omega) \left\{\frac{f(\omega)}{p_\rho
-\omega - i\eta} + \frac{f(-\omega)}{p_\rho -\omega+i\eta}\right\} ~,
\end{equation}
with
\begin{equation}
\label{I.2}
f(\omega) = (e^{\beta\omega}+1)^{-1} ~
{\stackrel{T=0}{_{\mbox{$\longrightarrow$}}}} ~
\Theta(-\omega) ~.
\end{equation}
The formal structure of the spectral function $A(\vec{p},\omega)$ \cite{15}
remains unaltered to the case for  $T=0$.

A further difference in comparison with Ref.\,\cite{12} is that we take the
momentum dependence of the self--energies into account. Since the pole of
the quasi--particle propagators occurs for $T\neq 0$ in the integration
domain of the intermediate states, one obtains, in principle, complex
effective scattering matrix elements and self--energies. It was checked in
Ref.\,\cite{12} that the imaginary part of $\Sigma$ turned out to be small.
Therefore we neglect also Im~$\Sigma$ in the calculations. For the
one--boson--exchange interaction we used the modern potentials constructed
by Brockmann and Machleidt \cite{22}. We select for the presentation  the
so-called potential $B$, which gives the best results for the EOS ($E/A =
-15.73$~MeV; $\rho_0 = 0.172$~fm$^{-3}; K = 249$~MeV; $J = 32.8$~MeV) at zero
temperature in RBHF--calculations (see Refs.\,\cite{15,15a}, for the
potential $A$ the outcome is similar, see Ref.\,\cite{21}). For the sake of
comparison we  also treated the RHF--approximation, where we adjusted the
force parameters to the outcome for the EOS for symmetric and asymmetric
nuclear matter at $T=0$ \cite{15a,21}. The RHF--approximation has in
comparison with the RH--approximation the advantage to resemble in its formal
structure more to the RBHF--approximation with the benefit of a much simpler
numerical treatment than in the RBHF case.

For finite temperatures one needs for the determination of the pressure the
free energy per baryon, defined as
\begin{equation}
\label{I.3}
f = u - T s ~,
\end{equation}
with the entropy per baryon:
\begin{equation}
\label{I.4}
s = - \frac{2}{\rho h^3} \sum_\tau \int d^3p\,[n(\vec{p})\; {\rm ln}
\,n(\vec{p}) +
  \left(1 - n(\vec{p}\right)\; {\rm ln}\,\left(1 - n(\vec{p}\right)]~.
\end{equation}
The pressure is given by:
\begin{equation}
\label{I.5}
P = \rho \sum_\tau \rho_\tau \left(\frac{\partial f}
    {\partial\rho_\tau}\right)_{T,\rho_{-\tau}} ~.
\end{equation}

\section{Results and discussion}
Common to almost all nonrelativistic treatments are EOSs typical in shape to
the standard Van de Waals behavior. The value for the critical temperature
depends strongly on the choice of the forces (and approximations). The
bandwidth reaches from 14 to 22~MeV, with a critical density of about 1/3 of
the saturation density $\rho_0$ \cite{1a}. In a recent calculation of
asymmetric matter for $T\neq 0$ a critical temperature of 20.8~MeV ($\rho_c =
0.39\,\rho_0$) was obtained for symmetric matter, which decreased to 8.0~MeV
($\rho_c = 0.24\,\rho_0$) for the limiting asymmetry of $\delta = 0.75$
\cite{2}. In the relativistic treatment within the framework of the
relativistic Hartree approximation the Van der Waals behavior is still
present, but in general one obtains lower critical temperatures (for
instance, $T_c = 14.4$~MeV, $\rho_c = 0.31\rho_0$ \cite{9}, $T_c \cong
14$~MeV \cite{10}).

Of great theoretical interest is the microscopic RBHF--treatment for two
reasons:   Firstly, as mentioned before, the RBHF--EOSs are much stiffer
than their nonrelativistic counterparts \cite{15b}.   Secondly, according to
Refs.\,\cite{1a,12} the typical Van der Waals behavior may be questionable,
and one obtains a rather low critical temperature, $T_c \simeq 8-9$~MeV
\cite{1a} or 12~MeV ($\rho_c\simeq 0.6\,\rho_0$) \cite{12}, depending on the
approximative treatment of the self--energy (see, also the discussion in
Ref.\,\cite{1a}). Due to these reasons we recalculated the
RBHF--approximation in the full Dirac space as described in
Refs.\,\cite{15,21} with the Brockmann--Machleidt potentials \cite{22}, which
give better results for the saturation properties \cite{15}. As discussed
before we treated also the RHF--approximation, where the force parameters are
adjusted to the outcome of the RBHF--treatment at $T=0$ \cite{15a,21}.

In Figs.\,1 and 2 we present
 first on a larger scale 
the EOSs for different asymmetries and temperatures,  computed for RBHF and
RHF, respectively. On this scale the energy per nucleon differs not very much
in both approximations. This holds also for the pressure (see Fig.\,3), free
energy and the entropy (see Fig.\,4). As expected the EOSs seem to be
stiffer as in non-relativistic Brueckner calculations (cf., for instance,
with Ref.\,\cite{5}). The entropy behavior is similar to Ref.\,\cite{5} and
agrees with the experimental situation (see, e.g., Fig.\,6 of
Ref.\,\cite{5}). 
Interesting is the EOS in the lower density domain, where one expects a
phase transition. The situation is depicted for symmetric matter in Figs.\,5
and 6, from which we deduce a critical temperature of 10.4~MeV (RBHF) and
15.2~MeV (RHF), respectively. (For asymmetric matter, see Ref.\,\cite{21}.)
The RBHF value is in accordance with Ref.\,\cite{12} and with one
nonrelativistic BHF\,--\,calculation \cite{5}. The RHF result for $T_c$ is
higher and agrees more or less with the result of Ref.\,\cite{9}, where one
uses a smaller equilibrium density.  The critical temperatures and pressures
for different asymmetries in the RBHF--treatment are given in Fig.\,7 (for
RHF, see Ref.\,\cite{21}). The critical densities are shown in Fig.\,8. For
symmetric matter they are slightly above 1/3 of the saturation density but
smaller than in Ref.\,\cite{12}.

One should however mention in this context that in general BHF--calculations
are facing numerical convergence problems for small densities \cite{5} and
may even not be applicable \cite{26}. For that reasons the conclusions drawn
from the BHF--approximation for low densities (and temperatures) should be
considered with some caution and a phenomenological description with
adjustable parameters, for instance, with Skyrme forces, might be a suitable
alternative in this sensitive domain. Unfortunately the critical temperature
can not be extracted directly from experiment and additional models and
assumptions are needed to obtain it from hot nuclei produced in heavy ion
collisions \cite{1a}. However more recent experiments  imply that a lower
critical temperature might be possible \cite{25}.

 As a final point we would like to mention that the symmetry energy remains
-- as expected -- a monotonic increasing function of the density in the case
for finite temperatures (see Fig.\,9), so that the composition of
proto-neutron stars should show, as in cold neutron stars for relativistic
EOSs, the tendency to lower asymmetries with increasing density \cite{23,13};
contrary to some nonrelativistic EOSs with a non-monotonic symmetry energy
\cite{24}.

\section{Conclusions}
In conclusion, we have performed a calculation of hot symmetric and
asymmetric nuclear matter within the relativistic Brueckner--Hartree--Fock
scheme using modern OBE--interactions constructed by Brockmann and Machleidt.
It turned out that the critical temperatures are smaller than it is the case
for the majority of nonrelativistic treatments. We have additionally treated
the relativistic Hartree--Fock approximation at $T\neq 0$, where the
Lagrangian parameters were adjusted to the outcome of the RBHF--treatment for
$T=0$. Here the critical temperature is in the range of other relativistic
treatments performed in the Hartree scheme.

\section*{Figure captions}
Fig.\,1: \hspace{3mm}\parbox[t]{12cm}{Energy per nucleon for different
asymmetries ($\delta = \frac{\rho_n - \rho_p}{\rho}$) and temperatures as
function of the density in the RBHF--approximation (Brockmann\,--\,Machleidt
potential B).}

Fig.\,2: \hspace{3mm}\parbox[t]{12cm}{Energy per nucleon for different
asymmetries and temperatures in the RHF\,--\,approximation.}

Fig.\,3: \hspace{3mm}\parbox[t]{12cm}{Density dependence of the pressure for
different asymmetries and temperatures in the RBHF\,--\,approximation. (In
the RHF\,--\,approximation the pressure increases less for higher
temperatures).}

Fig.\,4: \hspace{3mm}\parbox[t]{12cm}{Entropy per baryon for nuclear 
 matter versus density at different temperatures (for the experimental
comparison, see Fig.\,6 of Ref.\,\cite{5}).}

Fig.\,5: \hspace{3mm}\parbox[t]{12cm}{Pressure as a function of baryon
density for nuclear matter ($\delta = 0$) at different temperatures in the
RBHF\,--\,approximation.}

Fig.\,6: \hspace{3mm}\parbox[t]{12cm}{Pressure versus baryon density for
nuclear matter at different temperatures in the RHF\,--\,approximation.}

Fig.\,7: \hspace{3mm}\parbox[t]{12cm}{Critical temperature and pressure as
function of the asymmetry (RBHF; $T_c$ and $P_c$ are smaller than in the
RHF).}

Fig.\,8: \hspace{3mm}\parbox[t]{12cm}{Critical density as function of the
asymmetry.}

Fig.\,9: \hspace{3mm}\parbox[t]{12cm}{Symmetry energy as function of the
baryon density at different temperatures (RBHF).}

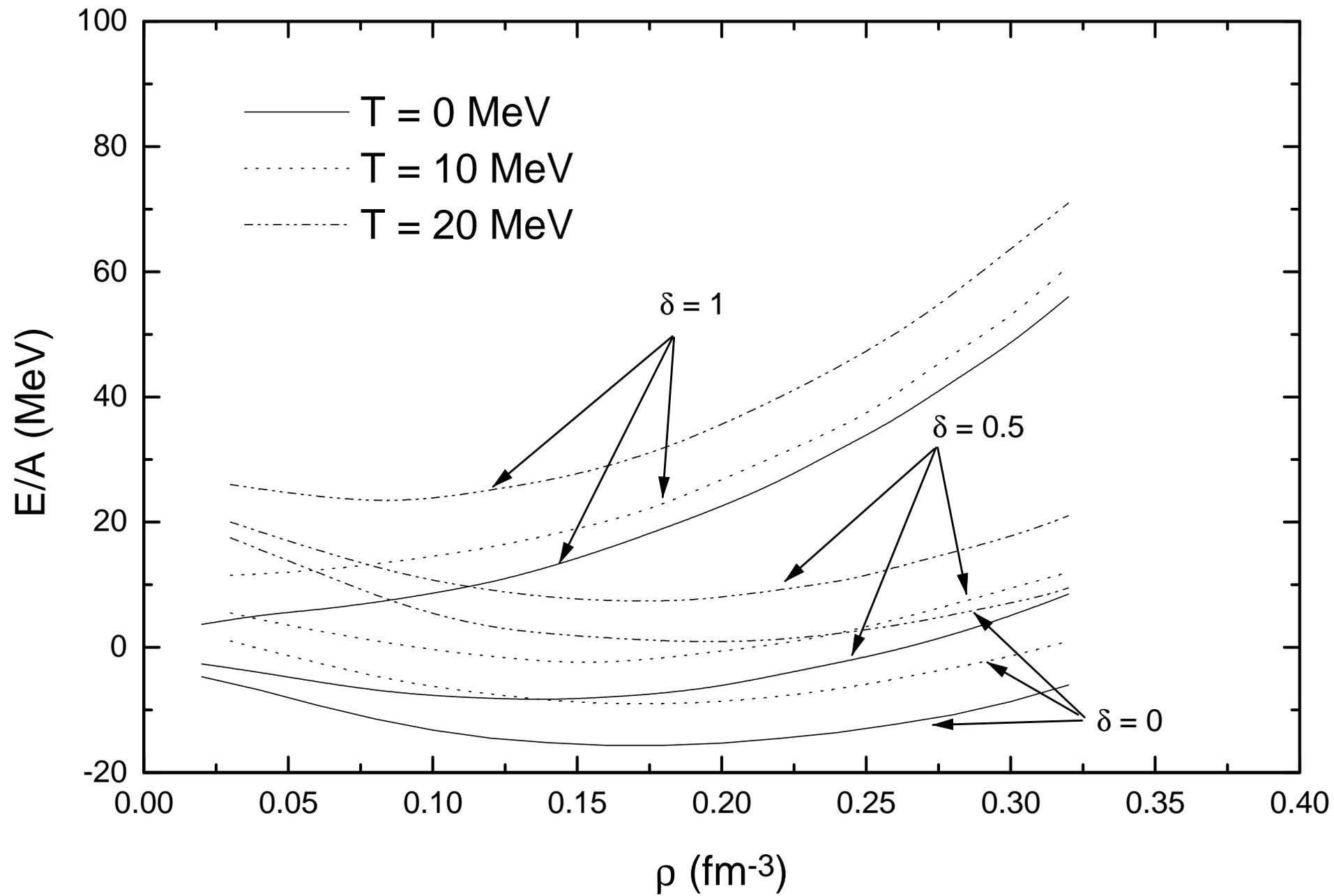
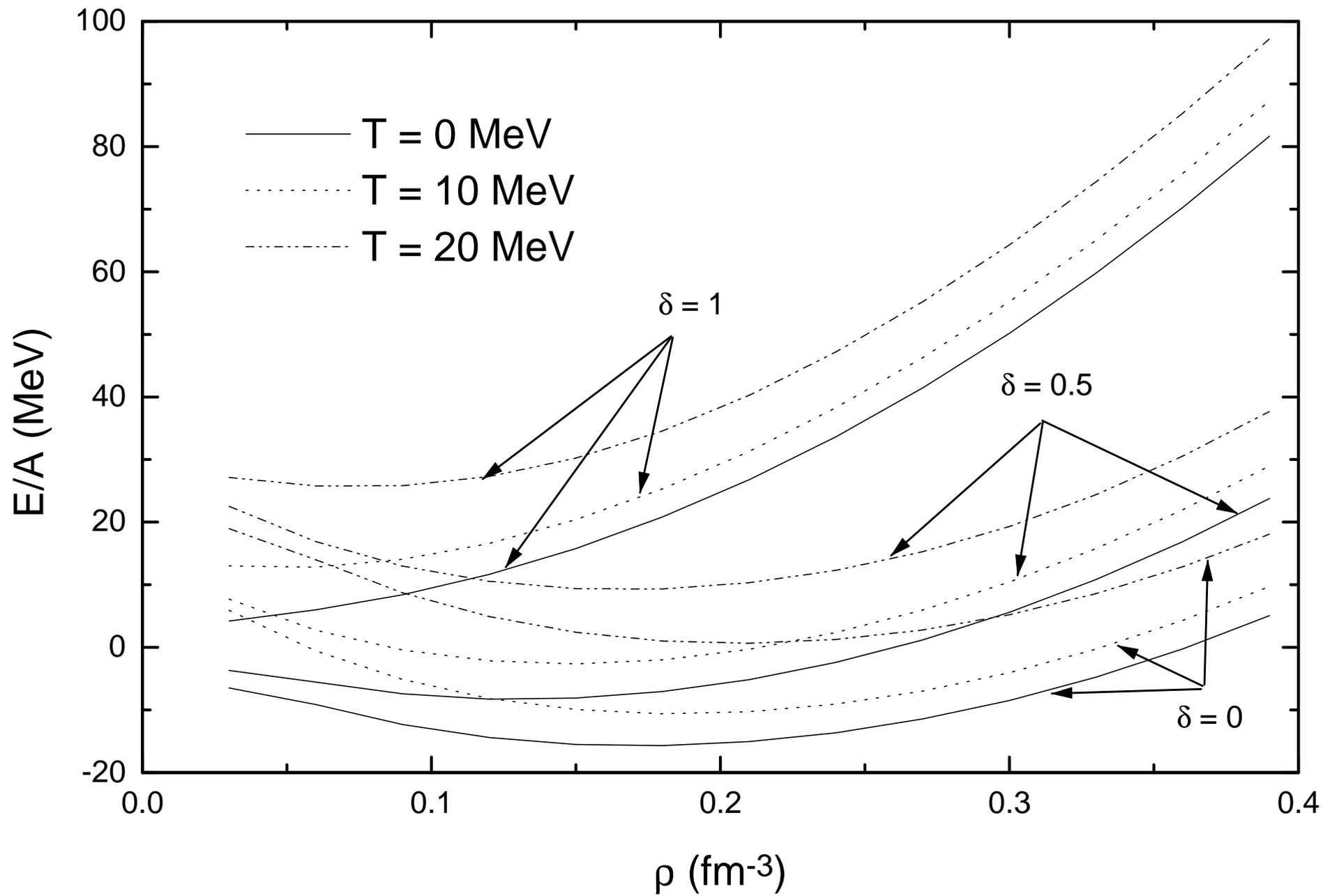
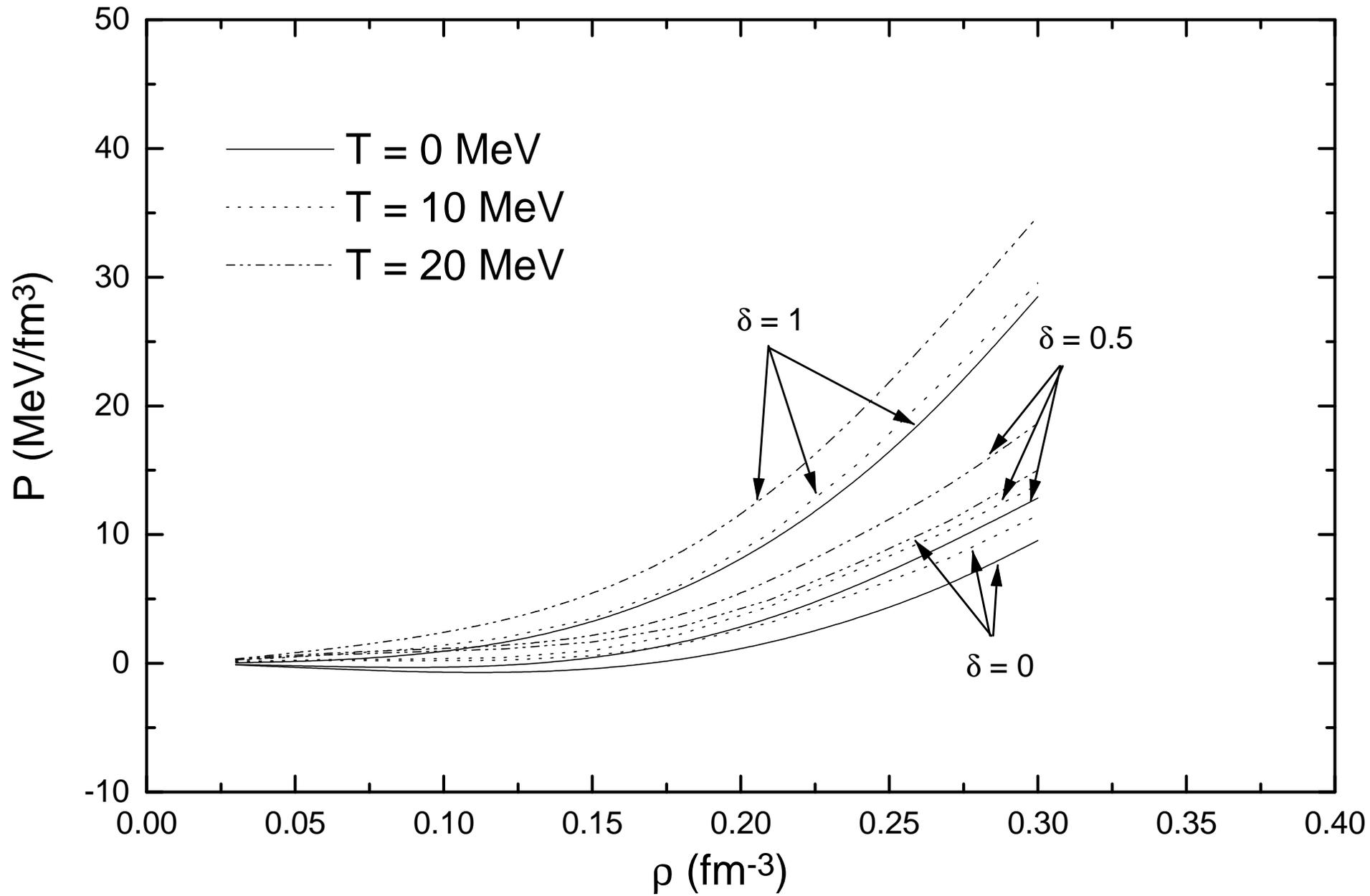
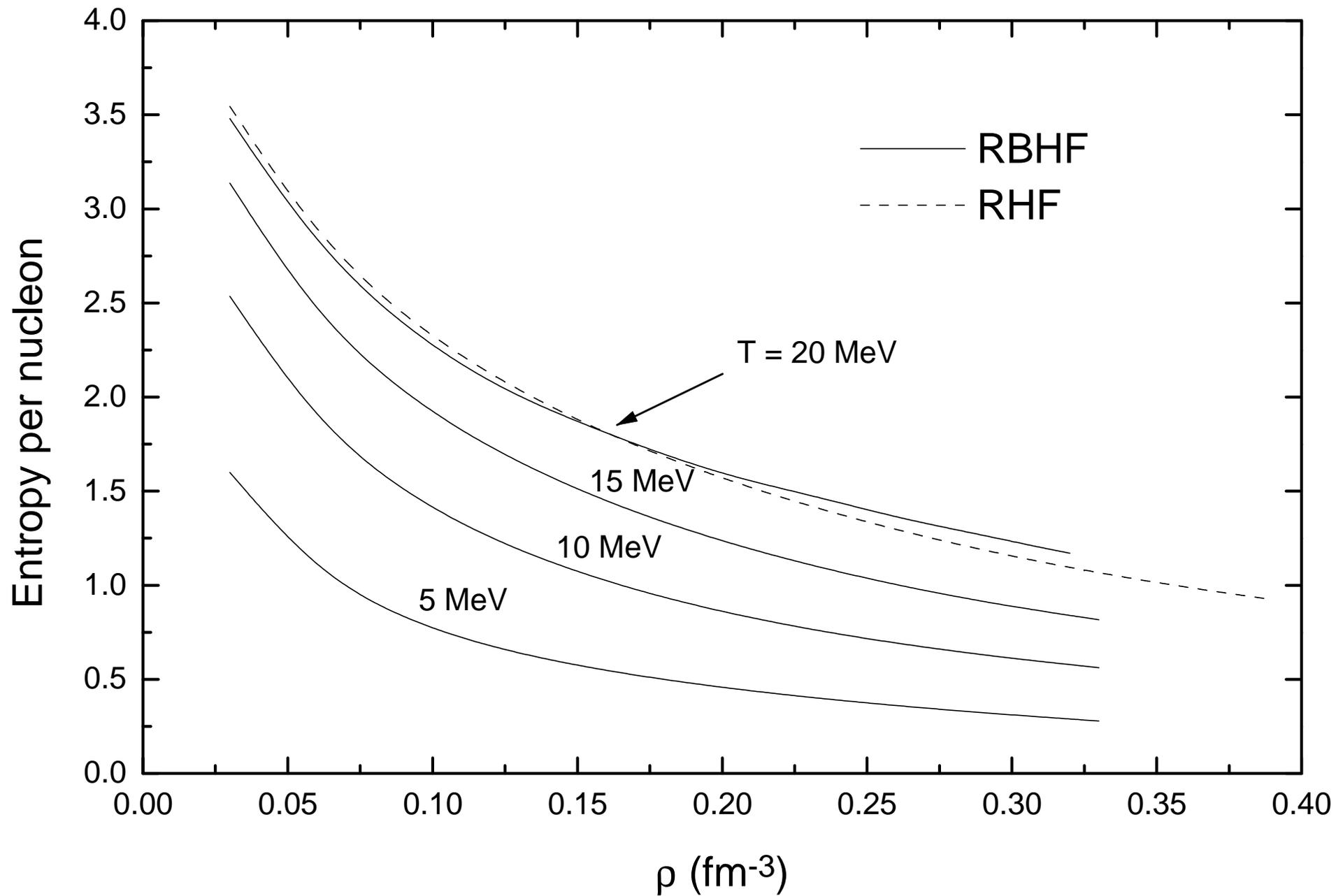
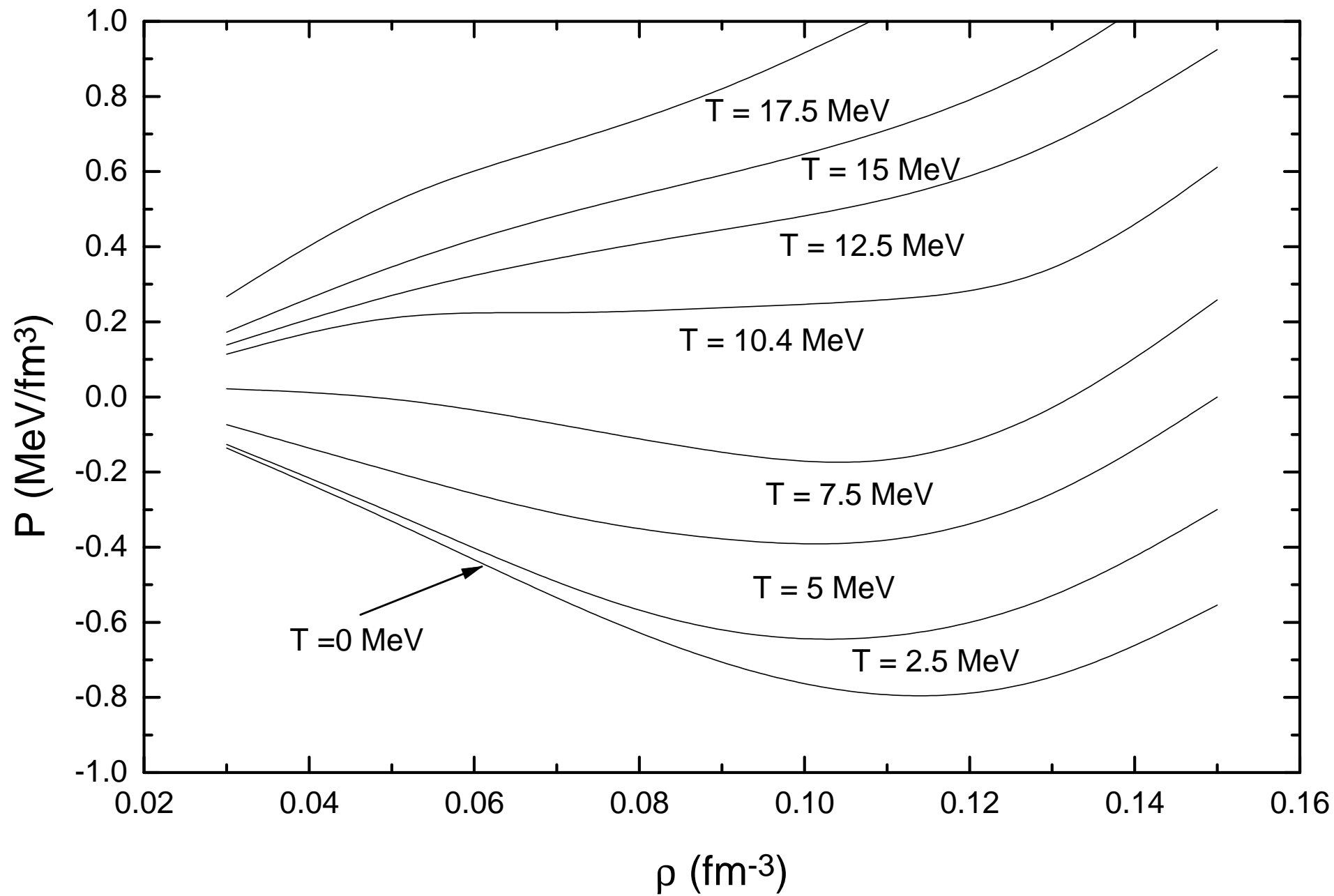
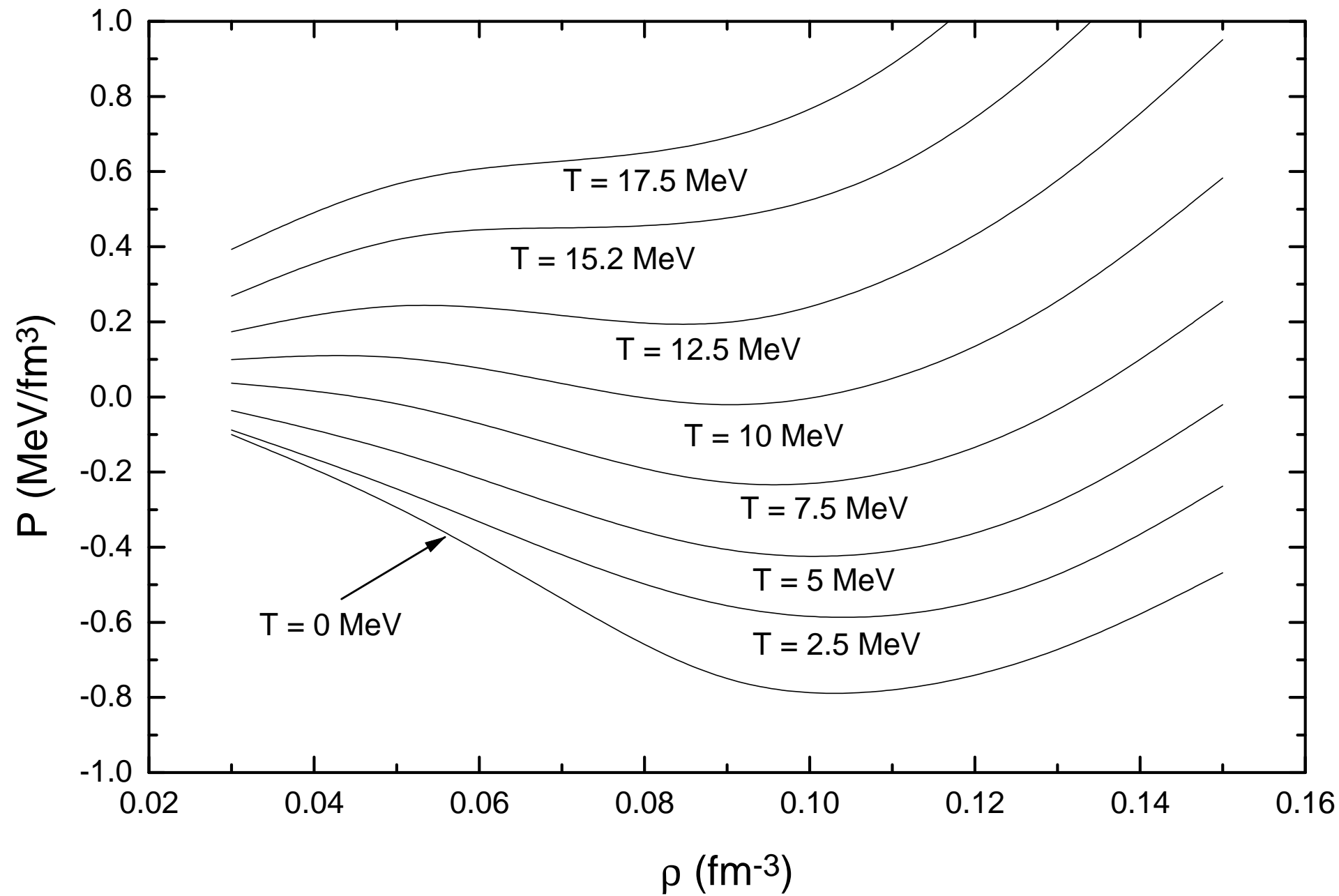
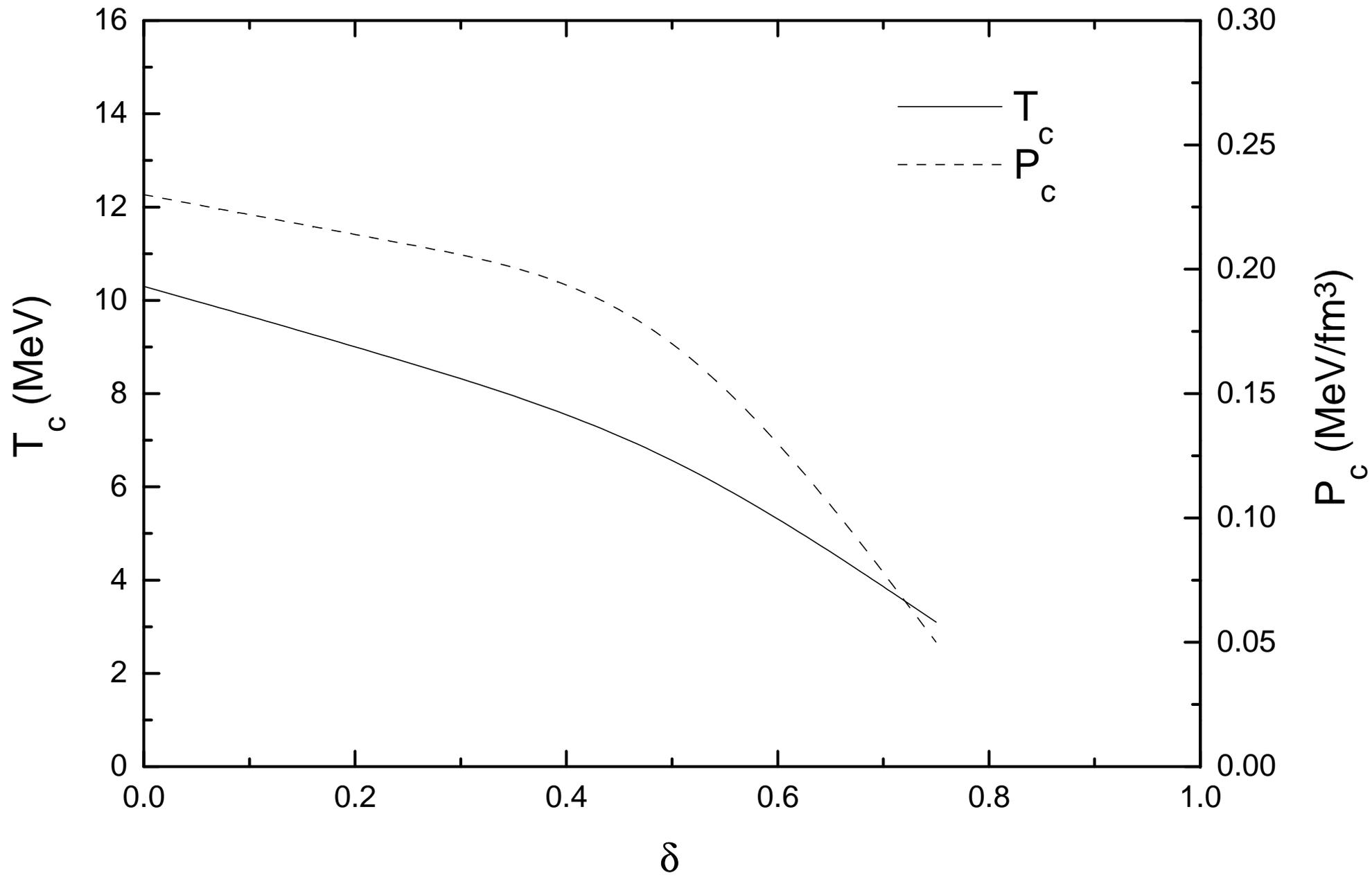
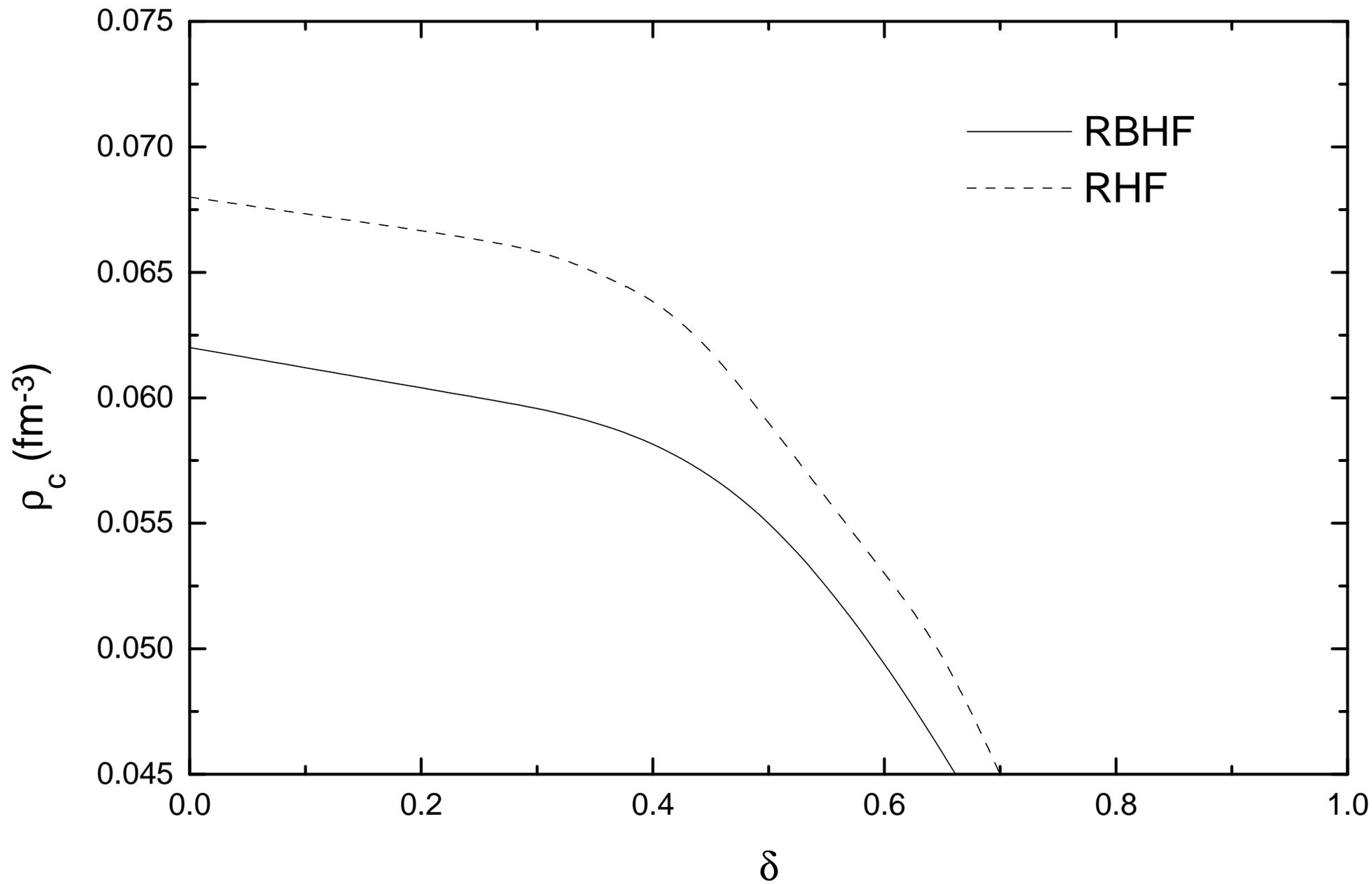
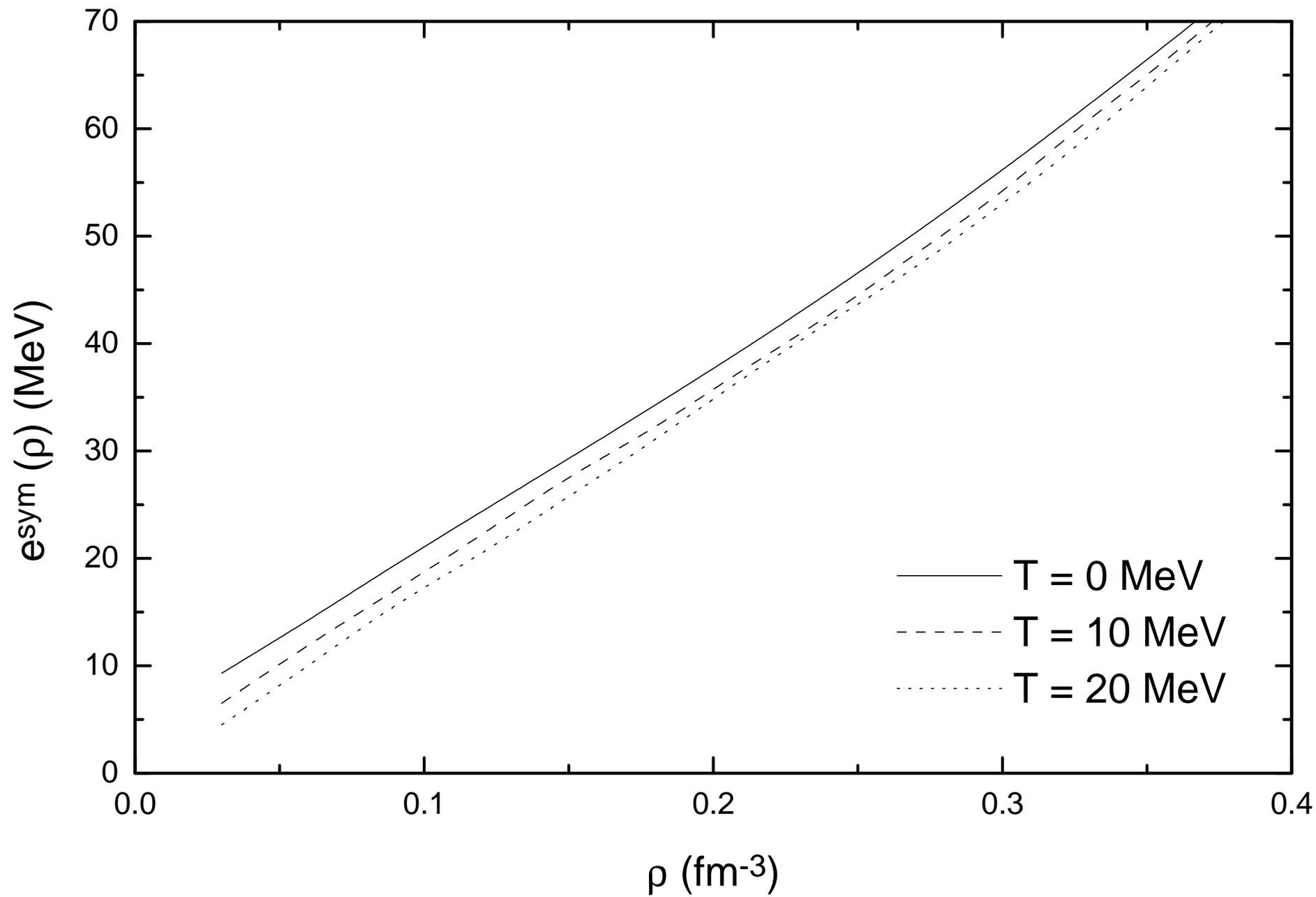
\end{document}